\documentclass[twocolumn,secnumarabic,amssymb, nobibnotes, aps, prc, superscriptaddress, nobalancelastpage,longbibliography]{revtex4-2}

 \usepackage{graphicx}
\usepackage{bm}
\usepackage{amsmath} 
\usepackage{braket} 
\usepackage{epsfig}
\usepackage{tensor}
\usepackage{CJKutf8}
\usepackage[T1]{fontenc}
\usepackage[utf8]{inputenc}
\usepackage{lmodern}
\usepackage[version=4]{mhchem}
\usepackage{ifthen}
\usepackage{listings}
\usepackage{array}
\usepackage{multirow}
\usepackage{floatrow}
\usepackage{enumitem}
\usepackage[colorlinks=true, allcolors=blue]{hyperref}
\usepackage[para,online,flushleft]{threeparttablex}
\usepackage{float}
\floatstyle{plaintop}
\restylefloat{table}

\usepackage{diagbox}
\usepackage{array}
\newlength{\colw}
\newcommand{\dc}[2]{\diagbox[width=\colw]{#1}{#2}}
\hypersetup{breaklinks=true,colorlinks=true,linkcolor=blue,citecolor=blue,filecolor=magenta,urlcolor=cyan}

\usepackage[all]{hypcap}

\usepackage{xcolor}
\definecolor{pastelgray}{rgb}{0.81, 0.81, 0.77}
\definecolor{beaublue}{rgb}{0.9, 0.9, 0.93}

\usepackage{color}

\usepackage{orcidlink}
\usepackage{tikz,xcolor}

\definecolor{lime}{HTML}{A6CE39}
\DeclareRobustCommand{\orcidicon}{
	\begin{tikzpicture}
	\draw[lime, fill=lime] (0,0) 
	circle [radius=0.16] 
	node[white] {{\fontfamily{qag}\selectfont \tiny ID}};
	\draw[white, fill=white] (-0.0625,0.095) 
	circle [radius=0.007];
	\end{tikzpicture}
	\hspace{-2mm}
}

\foreach \x in {A, ..., Z}{%
	\expandafter\xdef\csname orcid\x\endcsname{\noexpand\href{https://orcid.org/\csname orcidauthor\x\endcsname}{\noexpand\orcidicon}}
}

\makeatletter
\def\@bibdataout@aps{%
\immediate\write\@bibdataout{%
@CONTROL{%
apsrev41Control%
\longbibliography@sw{%
    ,author="08",editor="1",pages="1",title="0",year="1"%
    }{%
    ,author="08",editor="1",pages="1",title="",year="1"%
    }%
  }%
}%
\if@filesw \immediate \write \@auxout {\string \citation {apsrev41Control}}\fi
}
\makeatother

\newcolumntype{Y}{>{\centering\arraybackslash}X}

\usepackage[english]{babel}
\usepackage{booktabs}
\usepackage{tikz}

\begin{document}
\begin{CJK*}{UTF8}{gbsn}

\title{Unfolding of exotic near-threshold structure and decay dynamics in $^{17}$B}
\author{A. Volya\,\orcidlink{0000-0002-1765-6466}}\email{avolya@fsu.edu}
\affiliation{Department of Physics, Florida State University, Tallahassee, FL 32306, USA}

\author{S. M. Wang (王思敏)\,\orcidlink{0000-0002-8902-6842}}\email{wangsimin@fudan.edu.cn}
\affiliation{Key Laboratory of Nuclear Physics and Ion-beam Application (MOE), Institute of Modern Physics, Fudan University, Shanghai 200433, China}
\affiliation{Shanghai Research Center for Theoretical Nuclear Physics,
NSFC and Fudan University, Shanghai 200438, China}

\author{M. P{\l}oszajczak}\email{marek.ploszajczak@ganil.fr}
\affiliation{Grand Acc\'el\'erateur National d’Ions Lourds (GANIL), CEA/DSM and CNRS/IN2P3, BP 55027, F-14076 Caen Cedex, France}

\author{Z. C. Xu (许志成)\,\orcidlink{0000-0001-5418-2717}}\email{xuzhicheng@fudan.edu.cn}
\affiliation{Key Laboratory of Nuclear Physics and Ion-beam Application (MOE), Institute of Modern Physics, Fudan University, Shanghai 200433, China}
\affiliation{Shanghai Research Center for Theoretical Nuclear Physics,
NSFC and Fudan University, Shanghai 200438, China}

\begin{abstract}
Neutron-rich boron isotopes provide a valuable testing ground for threshold-driven structure and reaction phenomena, including halo formation and exotic decay modes. In particular, the structure of $^{17}\mathrm{B}$ and its relation to unbound $^{16}\mathrm{B}$ are of special interest. The $^{16}\mathrm{B}$ nucleus is slightly unbound by approximately 50~keV, while $^{17}\mathrm{B}$ is bound with a neutron separation energy of about 1.4--1.6~MeV. The observation of a 1640~keV $\gamma$ ray in $^{17}\mathrm{B}$, which we argue originates from a $1/2^{-}$ excited state, points to a remarkable situation in which $\gamma$ decay and two-neutron decay can compete.

We analyze 
and identify the main reasons for the competition: $L=2$ emission of the neutron pair, and structural realignment driven by the proximity of the one-body threshold, in particular the nearby $s$-wave neutron decay channel. 
The decay is a unique near-threshold $L=2$ process in which multiple structural components contribute, each with coexisting direct and virtual sequential amplitudes whose interference governs the observables.
Because threshold dynamics, continuum coupling, and interference of multiple quantum pathways are universal, closely related scenarios arise in ultracold atoms near Feshbach resonances, few-body atomic and molecular breakups, mesoscopic and photonic open systems, and other areas where open-quantum-system effects impact observables.

We employ advanced theoretical models to study this first-of-its-kind case and, 
provide a coherent theoretical perspective based on configuration interaction and complex-energy formalisms that incorporate both reaction continuum and structural effects near threshold. 
\end{abstract}

\maketitle

\end{CJK*}

\section{Introduction}
The physics of open quantum systems has emerged as a central theme in modern science, spanning diverse fields 
including atomic, molecular, optical, condensed matter, and nuclear physics~\cite{Rotter_2015}. Among these, nuclear physics stands 
out as one of the earliest and most extensively studied areas of quantum many-body theory, where openness due to 
coupling with the continuum is a defining feature~\cite{Mukha2006, Kohley2013, Kondo2023,Charity2023}. Atomic nuclei are natural open systems, and the foundational 
understanding of their structure and decay has deep roots in the pioneering works of Gamow, Wigner, and others, 
who introduced essential concepts such as quantum tunneling, resonance behavior, the dynamics of poles in the 
complex energy plane, clustering, and the role of the continuum and thresholds \cite{Gamow1928Zur, Wigner1948Behavior, Breit1957, Barker1964model, Michel2007, Michel2009, refId0, linaresfernandez:tel-04483932}.

An important consequence of openness is the emergence of strong threshold effects. In this regime, both structure and dynamics are often governed by the interference of multiple quantum pathways, including direct, virtual, resonant, sequential, and non-sequential processes, rather than by a single dominant path or decay mode. Such threshold-driven phenomena are universal and arise not only in nuclei but also in ultracold atomic systems near Feshbach resonances, few-body atomic and molecular breakups, and other open quantum systems where continuum coupling plays a decisive role~\cite{Feshbach1958Unified, Fano1961Effects, Kohler2006Production, Chin2010Feshbach}.

Recent experimental advances, particularly in radioactive beam facilities and precision spectroscopy, have brought these long-standing questions into new focus. They now allow direct access to weakly bound and unbound nuclear states, revitalizing the study of nuclear structure and reactions within the framework of open quantum systems.

In this work, we discuss the chain of exotic neutron-rich Boron isotopes. Boron, with $Z = 5$, has a single proton hole in the $p_{3/2}$ oscillator shell, which is closed at carbon. The chain of neutron rich boron isotopes starts with $^{13}$B which has eight neutrons and is often considered a closed neutron core, corresponding to a filled $p$-shell. 

In the $sd$-shell region, the relative positions of the $s_{1/2}$ and $d_{5/2}$ orbitals are of particular interest. On one hand, coupling to the continuum and weak binding significantly lower the $s$-orbital relative to the $d_{5/2}$ \cite{Hoffman2014Neutron}. On the other hand, as the number of neutrons increases, pairing correlations tend to favor filling orbitals with larger degeneracy. Additionally, the proton-neutron interaction between the proton hole and valence neutrons introduces an intriguing competition between structures involving the $s$- and $d$-wave configurations. The $s$ orbit for neutrons, where there is no centrifugal or Coulomb barrier, is most strongly affected near threshold, a transitional region between bound and virtual states, and is known to be associated with significant structural discontinuities. At the same time, as this orbit becomes fully occupied with an increased number of neutrons, its role transitions to an inert core, and it can no longer be part of reaction physics, since it is fully occupied in both the initial and final states.

\begin{figure*}[t]
    \centering
\includegraphics[width=0.8\linewidth]{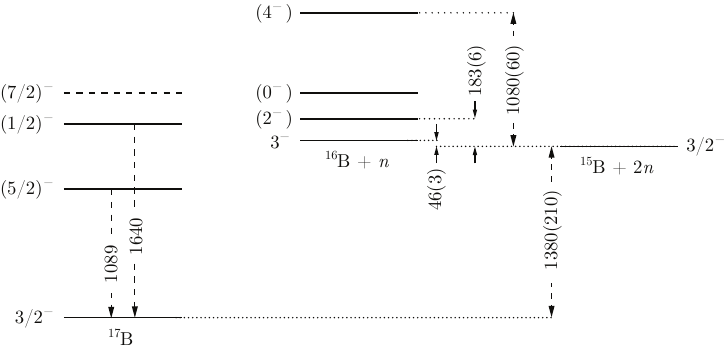}
\caption{Energy levels and transitions in boron isotopes in the region $^{15}\mathrm{B}$--$^{17}\mathrm{B}$. 
All energies are shown in keV. Levels are displayed relative to the $^{17}\mathrm{B}$ ground state, illustrating the $^{16}\mathrm{B}$ and $^{15}\mathrm{B}$ excitation spectra relative to the one- and two-neutron thresholds, respectively. 
Energy labels in $^{17}\mathrm{B}$ indicate excitation energies of the levels.
Two $\gamma$-ray energies from Ref.~\cite{Mozumdar2025Searcha} are indicated by dashed lines. 
For $^{17}\mathrm{B}$ and $^{16}\mathrm{B}$, spin assignments shown in parentheses are tentative from shell-model calculations and experiments; we present supporting evidence for these assignments in the text.
The dashed $7/2^{-}$ state in $^{17}\mathrm{B}$ reflects theoretical expectations, remains unobserved, and is included for illustration and discussion.
Low-lying states in $^{16}\mathrm{B}$ are taken from the combined experimental data of Refs.~\cite{Mozumdar2025Searcha,Yang2021Quasifree}.
Energies shown for states in $^{16}\mathrm{B}$ are resonance energies relative to the $^{15}\mathrm{B}$ ground state.
The two-neutron separation energy, $S_{2n}=1.38(21)\,\mathrm{MeV}$~\cite{and2012Ame2012}, is shown by a horizontal dotted line at the central value and corresponds to the $^{15}\mathrm{B}$ ground-state energy relative to $^{17}\mathrm{B}$; the comparatively large uncertainty is reflected in the quoted value.}
    \label{fig:energy_levels}
\end{figure*}

The nucleus $^{15}\mathrm{B}$, with two neutrons above $^{13}\mathrm{B}$, is known to exhibit a two-neutron halo structure~\cite{Kalpakchieva2000Spectroscopy, Kanungo2005, Kryger1996Upper, Marques2015Structure}. This closure of the $s$ shell makes the physics of the heavier $^{16}\mathrm{B}$ and $^{17}\mathrm{B}$ of considerable interest.
In Fig.~\ref{fig:energy_levels}, we illustrate the combined spectral information from the latest experimental data~\cite{Mozumdar2025Searcha, Yang2021Quasifree} for the $^{15}\mathrm{B}$–$^{17}\mathrm{B}$ region. 
Beyond the two-neutron $s$-shell closure, the addition of one neutron leads to a ground state in $^{16}\mathrm{B}$ that is unbound by approximately $40$--$100$~keV, as reported by multiple experiments, and several resonances have been observed. Their spin and parity assignments remain under discussion~\cite{Fortune2018Resonances,Mozumdar2025Searcha, Yang2021Quasifree}.
A neutron knockout experiment~\cite{Yang2015}, followed by proton removal experiment~\cite{Mozumdar2025Searcha} appear to have established a $3^-$ resonance at 46(3)~keV, which is interpreted as the ground state.
The experiment~\cite{Yang2015} did not observe a neutron knockout in the $s$-wave, but configuration interaction studies suggest that the corresponding spectroscopic factors are very small. This is consistent with the interpretation that the $(s_{1/2})^2$ configuration remains largely unchanged between the initial and final nuclei in the reaction process, thereby suppressing $s$-wave knockout strength.
Notably, the expected $0^-$ state was not observed, although many theoretical models predict it to lie even below the observed $3^-$ level. A $2^-$ state, discussed in Ref.~\cite{Fortune2018Resonances} and also predicted by theoretical models, has been suggested at 0.18~MeV of excitation ($\sim$64~keV from $^{15}\mathrm{B}+n$ threshold).

The next heavier $^{17}$B nucleus is well bound. Although its exact mass is not known, the neutron separation energy is believed to be around 1.4~MeV~\cite{ENSDF}. The ground state and the first excited state are established through the observation of $\gamma$-ray transitions, with $E_\gamma = 1.089$~MeV~\cite{Kanungo2005, Kondo2005Inbeam}. An additional $\gamma$ transition with 1640 keV
has been reported in~\cite{Gibelin2018Etats,Mozumdar2025Searcha}.
The spin and parity assignments for these levels in Fig. \ref{fig:energy_levels} 
are guided by the shell model; the higher spin $7/2^{-}$ state is also predicted in this region. 

The observed $\gamma$ ray at 1640~keV most likely originates from a state above the two-neutron decay threshold.
The excitation region in $^{17}\mathrm{B}$ around 1.4–1.6~MeV where the 1640 keV excited state is observed is particularly interesting because it lies near both the one- and two-neutron emission thresholds. In this window, one- and two-neutron decay channels can both contribute and can interfere through direct and (likely virtual) sequential pathways. 
The neutron decay of this 1640~keV state is also compelling from a theoretical perspective: the angular correlations in two-neutron emission are non-trivial, and near-threshold behavior for $L \ne 0$ decays remains largely unexplored. It is not clear whether any form of universality governs this kind of dynamics, or which spin and angular momentum couplings dominate in the two-neutron final state. Neutron correlations in this decay should be a sensitive probe of nuclear structure, in particular the previously discussed $s$-shell closure and the competition between $s$ and $d$ components.

 The observation of a $\gamma$ ray underscores the unique nature of this case, since neutron emission and electromagnetic decay compete even though their typical timescales are very different. 
 Understanding the apparent suppression of neutron emission, and what can be inferred from it, is therefore of interest. 
 One likely ingredient is the status of the one-body threshold, which may be closed. 
 Yet in this case the coupling to a low-lying virtual unbound state in $^{16}\mathrm{B}$, which could be $0^-,$ can influence the structure of the decaying state in $^{17}\mathrm{B}$ and its decay through the $\gamma$ and $2n$ channels. 
 Such effects of continuum coupling are known from other studies~\cite{Okolowicz2020Convenient,Volya2022Superradiance}. As we discuss in this work, this coupling may induce structural changes that further suppress direct two-neutron decay from $^{17}\mathrm{B}$ to $^{15}\mathrm{B}$, which could help explain the observed competition with the electromagnetic transition.

In this study, we examine this exotic near-threshold regime using structural calculations with both traditional configuration interaction shell model (SM) methods and the Gamow shell model (GSM)~\cite{Michel2002,Michel2009,Michel2021}, the latter incorporating continuum physics. We then analyze the two-neutron decay in detail through time dependent calculations. We assess the competition with $\gamma$ decay through a comparison of the relevant time scales, and investigate the role of a  one-body channel and the presence of a low-lying virtual state in the sequential decay path in their impact on the net $^{17}\mathrm{B}$ to $^{15}\mathrm{B}$ decay. Finally, we present predictions for neutron angular distributions, which may guide future experiments aimed at clarifying the structure and decay dynamics in this region.

\section{Structure of $^{17}\mathrm{B}$\label{sec:structure}}
In this section, we present a structural description of the states in $^{17}\mathrm{B}$, together with spectroscopic information on their configurations and decay channels, using a configuration-interaction framework that sets the stage for a quantitative analysis.
As shown in Fig.~\ref{fig:energy_levels}, the ground state in $^{17}\mathrm{B}$ has spin–parity $3/2^{-}$, which is broadly accepted. The excited states predicted by configuration–interaction calculations place $5/2^{-}$ first, followed by close-lying $1/2^{-}$ and $7/2^{-}$. These predictions for the low-lying states are relatively consistent among typical semi-empirical theoretical models, including the YSOX Hamiltonian~\cite{Yuan2012Shellmodel} and the more recent FSU interaction Hamiltonian~\cite{Lubna2019Structure, Lubna2020Evolution}. 

Experimentally, as stated earlier, the excited states have been identified through two observed $\gamma$ 
rays at 1089 and 1640~keV. The 1089~keV $\gamma$ was measured using in-beam $\gamma$-ray spectroscopy~\cite{Kanungo2005, Kondo2005Inbeam}; a detailed DWBA analysis supports the $5/2^{-}$ assignment for this first excited state. The 1640~keV state was recently observed in a proton-removal reaction from $^{18}\mathrm{C}$~\cite{Gibelin2018Etats, Mozumdar2025Searcha}; in that case, the 1089~keV $\gamma$ was not seen. From a structural standpoint, boron isotopes have one proton hole in the $p$ shell, which includes the $p_{1/2}$ and $p_{3/2}$ orbits from which the proton must be removed going from carbon; consequently, only the $3/2^{-}$ and $1/2^{-}$ states are expected to be directly populated in proton removal reaction. 
This is fully supported by the SM studies. 
The non-observation of the 1089~keV transition in this reaction further supports the $5/2^{-}$ assignment for the first excited state; this also excludes $7/2^{-}$ state as a source of the observed 1640-keV $\gamma$. 
The YSOX SM suggests proton-removal spectroscopic factors of 2.5 and 0.2 for the $3/2^{-}$ and $1/2^{-}$ states, respectively.

Since the $1/2^{-}$ and $7/2^{-}$ states lie close in the SM results, their ordering may be uncertain; however, placing the $7/2^{-}$ below the established $1/2^{-}$ would make it bound and thus $\gamma$-decaying, which is not supported by experiment.

Within the semi-empirical SM framework, we restrict our presentation to the YSOX Hamiltonian, since alternative interactions yield comparable spectra; YSOX has also been widely used in experimental analyses~\cite{Mozumdar2025Searcha}. In addition, the weak binding of this system motivates the use of the GSM, which includes continuum coupling and provides a more complete description of structure and decay near threshold.
In this case, the valence effective Hamiltonian is obtained via many-body perturbation theory (MBPT) with chiral interaction~\cite{Hu2020,Xu2023}, further details can be found in Appendix~\ref{sec:GSM}.

The two approaches are complementary. The semi-empirical Hamiltonians are well established and provide reliable, broadly predictive descriptions across many nuclei without additional adjustments, whereas the GSM offers a more advanced treatment of continuum coupling that can access the physics of interest here, albeit with slight fine-tuning to this specific case.
For electromagnetic transitions, we employ effective charges $e_p=1.5$ and $e_n=0.5$ for protons and neutrons, respectively within both models. 

\begin{table}[htbp]
\centering
\caption{Electromagnetic reduced transition probabilities $B(M1)$ (in $\mu_N^2$) and $B(E2)$ (in $e^2\text{fm}^4$) between the lowest states in $^{17}$B,
calculated with SM (YSOX interaction) and GSM (chiral effective interaction). The $\gamma$-ray energies ($E_\gamma$) are listed for the assumed transitions. 
Partial widths that assume theoretical reduced transition probabilities and experimental gamma energies are given in the last column in MeV.}
\begin{tabular}{ccccccc}
\toprule
Initial $\rightarrow$ Final & Type & $B^{\rm SM}$ & $B^{\rm GSM}$ & $E_\gamma$ (MeV) & $\Gamma$ (MeV) \\
\midrule
$5/2^-_1 \rightarrow 3/2^-_1$ & M1 & 0.452 & 0.732 & 1.089 & $6.76 \times 10^{-9}$ \\
                              & E2 & 22.714 & 46.253 & 1.089 & $2.81 \times 10^{-11}$ \\
\midrule
$1/2^-_1 \rightarrow 3/2^-_1$ & M1 & 0.305 & 0.227 & 1.640 & $1.56 \times 10^{-8}$ \\
                              & E2 & 10.287 & 4.220 & 1.640 & $9.84 \times 10^{-11}$ \\
$1/2^-_1 \rightarrow 5/2^-_1$ & E2 & 0.791 & 0.723 & 0.551 & $3.24 \times 10^{-14}$ \\
\midrule
$7/2^-_1 \rightarrow 3/2^-_1$ & E2 & 9.174 & 2.105 & 1.640 & $8.78 \times 10^{-11}$ \\
$7/2^-_1 \rightarrow 5/2^-_1$ & M1 & 1.037 & 0.616 & 0.551 & $5.29 \times 10^{-8}$ \\
                              & E2 & 20.165 & 0.342 & 0.551 & $8.26 \times 10^{-13}$ \\
\bottomrule
\end{tabular}
\label{tab:17B_gamma_widths}
\end{table}

In Table~\ref{tab:17B_gamma_widths}, the predicted $\gamma$-decay reduced transition probabilities and widths are presented. The widths are computed from the reduced transition probabilities for $M1$ and $E2$ transitions obtained in the SM and GSM calculations and listed in the table; the phase-space factors use the experimental $\gamma$-ray energies. As illustrated in Fig.~\ref{fig:energy_levels}, only two $\gamma$ rays, 1089 and 1640~keV, have been observed, and the sequential decay of the 1640~keV state (first a 551~keV $\gamma$, followed by a 1089~keV $\gamma$) has not been seen. In the table, we consider both possibilities for the 1640~keV transition, namely that it originates from the $1/2^{-}$ state or from the $7/2^{-}$ state.

As is typical for light nuclei, $M1$ transitions are several orders of magnitude faster than $E2$, which generally favors competition with neutron decay. 
It is therefore likely that the 1640~keV state has $J^\pi=1/2^{-}$. The $7/2^{-}$ level can decay to the $3/2^{-}$ ground state only via $E2$; furthermore,  $7/2^{-}$ state would be expected to decay sequentially, since the $7/2^{-}\to 5/2^{-}$ transition at 551~keV is $M1$ and much faster than a direct $E2$ branch to the ground state. 
However, the sequential $\gamma$ cascade (551~keV followed by 1089~keV) has not been observed. 
This absence is naturally explained by a $1/2^{-}$ assignment, for which the branch to the intermediate $5/2^{-}$ level is $E2$ and suppressed by nearly six orders of magnitude as seen in Table~\ref{tab:17B_gamma_widths}, so it is not expected to be seen.


\subsection{Direct pair removal\label{sec:pair}}

Following the above analysis, we conclude that the second excited state, $1/2^{-}$ in $^{17}\mathrm{B}$, decays to the ground state via the 1640~keV $\gamma$ ray. With the previously estimated two-neutron threshold at 1.38(21) MeV, \cite{and2012Ame2012}; this places the state approximately one standard deviation above the two-neutron separation threshold. As noted earlier, the one-neutron threshold lies 40–100~keV higher (recent work in Ref.~\cite{Yang2021Quasifree} suggests 46(3)~keV). This raises the possibility that the one-neutron decay channel is closed, while the two-neutron channel is open. 

Since the ground state of $^{15}\mathrm{B}$ is $3/2^{-}$, a direct two-neutron decay from the $1/2^{-}$ state in $^{17}\mathrm{B}$ requires an $L=2$ transfer. This higher orbital angular momentum, together with the much slower phase-space scaling at low energies~\cite{Volya2012Physics}, reduces the two-neutron decay rate and opens the possibility of competition with electromagnetic decay.

SM calculations of pair removal from $^{17}$B($1/2^-$) to the ground state of $^{15}$B ($3/2^-$) suggest a spectroscopic factor of approximately 0.6. The transferred neutron pair consists primarily of two configurations: $(0d_{5/2})^2$, contributing about 55\%, and a mixed configuration $(0d_{5/2})^1(1s_{1/2})^1$, contributing roughly 38\%. For brevity, in what follows we denote these configurations as $d^2$ and $sd$. 
The pair is in a $J=2$ state, with the spin part strongly dominated by the singlet $S=0$ component. 

While the SM spectrum is insensitive to continuum coupling, which depends on the precise positions of levels and on virtual excitations into the one-body continuum (topics discussed separately), it nevertheless captures the magnitude of the spectroscopic factor that arise from the many-body structure. 
The result also narrows the discussion of decay dynamics to the interplay between the configurations of emitted neutron pair $d^{2}$ and $sd$, both with spin $S=0$ and total angular momentum $J=2$.
As noted in the introduction, this mixing is expected to be modified by continuum coupling and near-threshold dynamics. The associated interference is of particular interest for correlation studies, since it may reveal characteristic patterns that help isolate the $s$-wave component closely tied to the halo character of boron isotopes.

\subsection{Decay path via resonances in \texorpdfstring{$^{16}$B}{16B}}

Whether the one-body decay channel is open or closed, the $1/2^{-}$ state lies either above this threshold or less than a few tens of keV below it; in both cases, strong threshold effects are expected, especially in the $s$-wave neutron channel $^{16}\mathrm{B}+\nu(s_{1/2})$. To couple to the $1/2^{-}$ state in $^{17}\mathrm{B}$, the state in $^{16}\mathrm{B}$ must have $J^\pi=0^{-}$ or $1^{-}$. This naturally leads to a discussion of the states in $^{16}\mathrm{B}$, which we take up next.

The structure of the ground and low-lying states of $^{16}$B has been a subject of active discussion. 
It is now established from multiple experiments that $^{16}$B has no bound states. 
Bowman \textit{et al.}~\cite{Bowman1974Detection} confirmed earlier suggestions~\cite{Kryger1996Upper} that all states in $^{16}$B are unbound. 
Numerous resonances have since been identified, with the most comprehensive survey provided in Refs.~\cite{Mozumdar2025Searcha,Yang2021Quasifree}, where the $^{17}$B$(p,pn)$ and $^{17}$B$(p,2p)$ reactions were used to populate excited states in $^{16}$B. 

In Table~\ref{tab:sequential_decay_transposed}, we list low-lying states in $^{16}\mathrm{B}$ that could play a role in the sequential decay process, along with spectroscopic factors for the associated transitions. We also include spectroscopic factors from the $(p,pn)$ and $(p,2p)$ reactions that have been used to populate these states \cite{Mozumdar2025Searcha,Yang2021Quasifree}. 
For the most part, the SM and GSM results agree in their qualitative description. In making this comparison, we note that spectroscopic factors are not genuine observables and are strictly well defined only in the perturbative, weak continuum-coupling limit. Consequently, their values can differ substantially between models with and without explicit continuum coupling. Nevertheless, both approaches consistently identify the same low-lying states and their role in the processes of interest.
The main exception is the $2^{-}$ state, where both models predict another nearby $2^{-}$ level that can mix with the lowest one. Since the $2^{-}$ state does not affect our discussion, and we show only the lowest $2^{-}$, we do not investigate this mixing further.

\begin{table}[h!]
\centering
\caption{Spectroscopic factors for the sequential decay path $^{17}\mathrm{B}$ ($1/2^-$) $\to$ $^{16}\mathrm{B}$ $\to$ $^{15}\mathrm{B}$ ($3/2^-$). Each cell reports the SM (lower-left) and GSM (upper-right) values. Columns list candidate intermediate states in $^{16}\mathrm{B}$. The first row gives experimental excitation energies, informed by recent data and our interpretation in the text. Only the lowest $3^-$ state is relatively well established; all other assignments are tentative and some are debated. Subsequent rows show $s$- and $d$-wave spectroscopic factors for transitions from the $^{17}\mathrm{B}$ $1/2^-$ state to these $^{16}\mathrm{B}$ states; the next set of rows shows transitions from these states to the $^{15}\mathrm{B}$ ground state ($3/2^-$). The final rows compile spectroscopic factors for reactions that populate these states \cite{Mozumdar2025Searcha,Yang2021Quasifree}.}
\begin{tabular}{|l|c|c|c|c|c|}
\hline
 & $0^-$ & $3^-$ & $2^-$ & $1^-$ & $4^-$ \\
\hline\hline
$E_x$ (MeV)
 & \dc{1.48}{0.29}
 & \dc{0.00}{0.00}
 & \dc{0.14}{0.38}
 & \dc{2.75}{1.72}
 & \dc{1.03}{1.27} \\
\hline
$s:{}^{17}\mathrm{B}\rightarrow{}^{16}\mathrm{B}$
 & \dc{0.34}{0.69}
 & --
 & --
 & \dc{0.21}{0.04}
 & -- \\
\hline
$d:{}^{17}\mathrm{B}\rightarrow{}^{16}\mathrm{B}$
 & --
 & \dc{0.11}{0.33}
 & \dc{0.31}{0.47}
 & \dc{0.03}{0.00}
 & -- \\
\hline\hline
$s:{}^{16}\mathrm{B}\rightarrow{}^{15}\mathrm{B}$
 & --
 & --
 & \dc{0.00}{0.58}
 & \dc{0.18}{0.54}
 & -- \\
\hline
$d:{}^{16}\mathrm{B}\rightarrow{}^{15}\mathrm{B}$
 & \dc{0.08}{0.08}
 & \dc{0.42}{0.53}
 & \dc{0.60}{0.40}
 & \dc{0.51}{0.27}
 & \dc{0.74}{0.74} \\
\hline\hline
$s:{}^{17}\mathrm{B}(\mathrm{p,pn})$
 & --
 & --
 & \dc{0.07}{0.15}
 & \dc{0.04}{0.04}
 & -- \\
\hline
$d:{}^{17}\mathrm{B}(\mathrm{p,pn})$
 & \dc{0.01}{0.02}
 & \dc{0.86}{0.75}
 & \dc{0.60}{0.10}
 & \dc{0.12}{0.04}
 & \dc{0.68}{0.51} \\
\hline\hline
$p:{}^{17}\mathrm{C}(\mathrm{p,2p})$
 & \dc{0.22}{0.95}
 & \dc{0.76}{0.40}
 & \dc{0.28}{0.02}
 & \dc{0.37}{0.31}
 & -- \\
\hline
\end{tabular}

\label{tab:sequential_decay_transposed}
\end{table}

A narrow resonance near threshold, with a width $\Gamma < 20$~keV, was observed following proton removal from $^{17}$C in Ref.~\cite{Lecouey2009Singleprotona}. The small width suggests $\ell = 2$ for the decay neutron. The small width suggests an $\ell = 2$ decay for the emitted neutron. As argued in Ref.~\cite{Lecouey2009Singleprotona}, considering that the ground state of $^{17}$C has $J^\pi = 3/2^+$ and that a $p_{3/2}$ proton is removed, the populated state in $^{16}$B can have $J^\pi = 0^-$, $1^-$, $2^-$, or $3^-$. The absence of an $s$-wave decay component for the decay of this state to the ground state of $^{15}$B implies that the populated $^{16}$B state is most likely either $J^\pi = 0^-$ or $3^-$. The state was located at $85(15)$~keV above threshold, but estimates range, other quoted results include $40(40)$~keV and $60(20)$~keV. However, in the $^{17}$B$(p,pn)$ reaction of Ref.~\cite{Yang2021Quasifree}, this same low-lying resonance was also populated, which indicates that it is more likely the $3^-$ state, since the spectroscopic factor for a $0^-$ state 
in this reaction is very small, as shown in Table~\ref{tab:sequential_decay_transposed}. Recently this ground state resonance was populated also by $(p,2p)$ reaction \cite{Mozumdar2025Searcha}. Taken together, these measurements now provide a more definite picture: the ground-state resonance of $^{16}$B is $J^\pi=3^-$ at an energy of about 46~keV above threshold.

Fortune~\cite{Fortune2018Resonances} argued that the ground state of $^{16}\mathrm{B}$ is $2^-$, citing the expected lowering of the $s_{1/2}$ orbital and a dominant configuration of three neutrons in the $sd$ shell forming a $1/2^+$ subsystem (rather than $3/2^+$). Coupling this subsystem to a $p_{3/2}$ proton hole yields a negative-parity multiplet in which the $2^-$ member is expected to lie lowest. For context, the ground state of $^{18}\mathrm{B}$ is $2^-$. However, the $2^-$ state is not established as the lowest in $^{16}\mathrm{B}$; nonetheless, Ref.~\cite{Yang2021Quasifree} assigned the level at $E_x=0.137$~MeV as $2^-$.

Many theoretical treatments predict the $0^-$ state to be the ground state of $^{16}\mathrm{B}$. 
The $0^-$ level is a member of the multiplet generated by coupling $^{17}\mathrm{C}(3/2^+)\!\otimes\!(p_{3/2})^{-1}$. Several other members of this multiplet have been identified experimentally, so the $0^-$ state is expected at low excitation energy.
The spectroscopic factor for the transition from the $0^-_1$ state in $^{16}\mathrm{B}$ to the $3/2^-$ ground state of $^{15}\mathrm{B}$ is small  as it involves neutron removal from the $d_{3/2}$ orbital. By contrast, the spectroscopic factor to the first excited $5/2^-$ state in $^{15}\mathrm{B}$ is large, so the $0^-$ state should decay preferentially to $5/2^-$ if energetically allowed. 
This state is not expected to be populated in the $(p,pn)$ reaction because the relevant spectroscopic factor is small (see Table~\ref{tab:sequential_decay_transposed}), but it should be observable in $(p,2p)$. An analysis of the results of Ref.~\cite{Mozumdar2025Searcha} suggests a candidate at $1.48$~MeV (excitation energy $E_x$) consistent with these expectations.

The last intermediate state of interest is the $1^{-}$ level; in Ref.~\cite{Yang2021Quasifree}, the 2.75~MeV state was interpreted as either $1^{-}$ or $2^{-}$. Both $0^-$ and $1^-$ provide strong $s$-wave coupling to the $1/2^-$ state in $^{17}\mathrm{B}$, leading to a unique near-threshold configuration. 
It is likely that the $1/2^{-}$ state lies below both the $0^{-}$ and $1^{-}$ resonances in $^{16}\mathrm{B}$, which makes any sequential process through these states of significantly reduced probability; the decay therefore can only proceed through their virtual excitation and via tails of these resonances. 
Nevertheless, the $1/2^-$ state is strongly coupled to these $s$-wave thresholds, and its structure is expected to be significantly influenced by the thresholds proximity \cite{Okolowicz2020Convenient}.

\section{Decay of $^{17}\mathrm{B}$ in the Gamow Coupled-Channel Framework}
From the analysis described in the previous section, one can conclude the following. In $^{17}\mathrm{B}$, the $1/2^{-}$ state at around 1640~keV of excitation lies in the direct vicinity of the one- and two-neutron decay thresholds. 
Since $^{16}\mathrm{B}$ is unbound, the one-body threshold is higher, thus decay via these resonances could only contribute as a simultaneous two-neutron emission process. 
Although there is no exact mass measurement, previous experimental studies place the two-neutron threshold at 1.38(21) MeV in $^{17}\mathrm{B}$, which clearly makes the $\gamma$-decaying $1/2^{-}$ state two-neutron unbound.

From the theoretical perspective, several important factors may be at play. 
While the direct contribution via the $s$-wave resonances $0^{-}$ and $1^{-}$ in $^{16}\mathrm{B}$ may be small, near-threshold effects that lead to structural changes of the $1/2^{-}$ state are expected to be significant and can hinder direct two-neutron decay. 
The two-neutron decay is expected to proceed with total angular momentum $J=2$, which by itself lowers the decay probability. 
Furthermore, the state of the emitted two neutrons may involve different configurations that are all kinematically possible at very low energies. 
This contrasts with the typical $J=0$ emission, where the low-energy limit is generally universal~\cite{Volya2012Physics}. 
The presence and interference of different configurations in the decay result in specific patterns in the observable two-neutron correlations and may provide an experimental handle for studying this physics. 

\begin{figure}[htb]
\centering
\includegraphics[width=0.9\columnwidth]{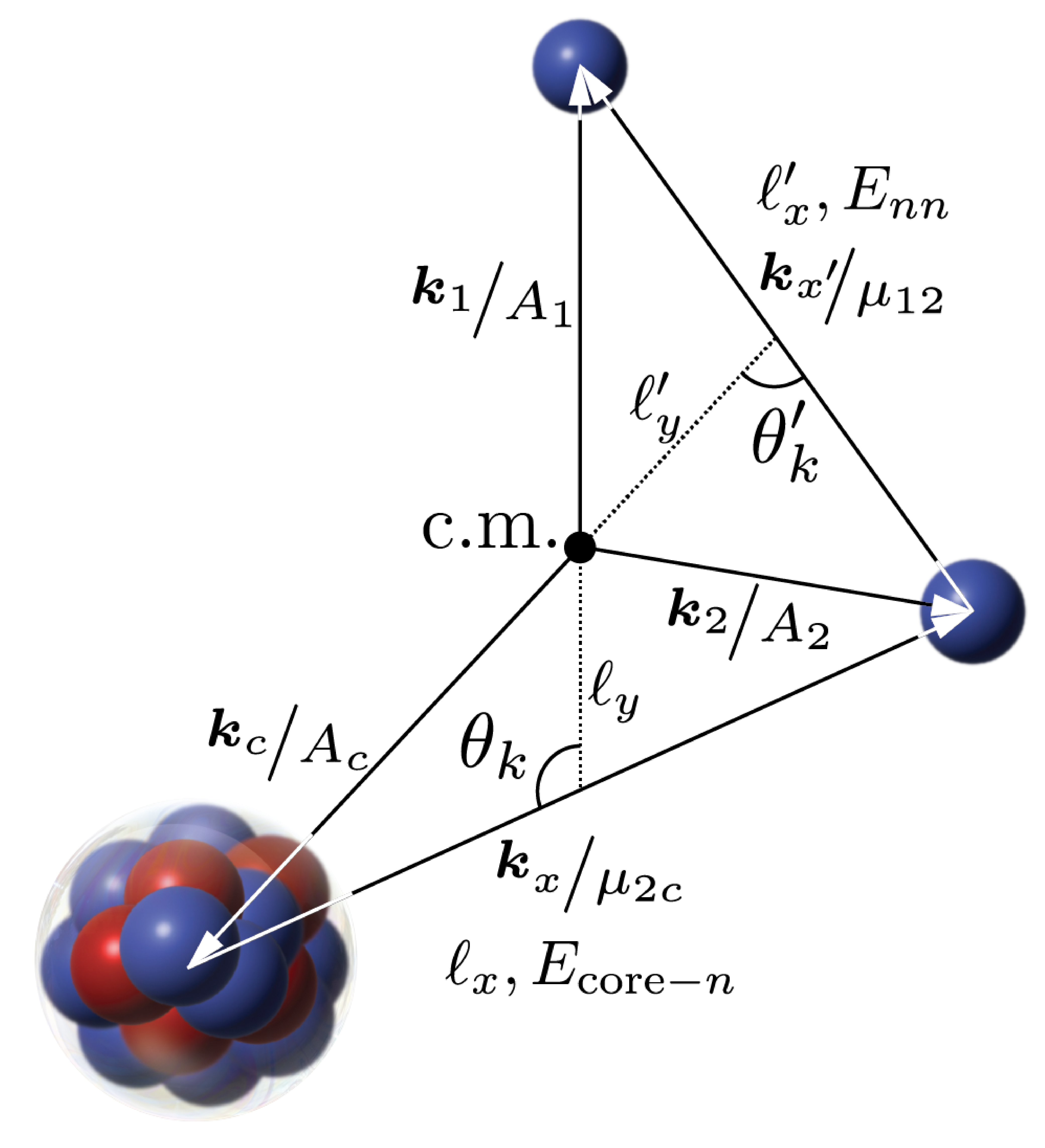}
\caption{Momentum-space kinematics for the three-body system $^{15}\mathrm{B}+n+n$. The diagram shows the center of mass (c.m.) and the single-particle momenta $\boldsymbol{k}_1$, $\boldsymbol{k}_2$, and $\boldsymbol{k}_c$ of the two neutrons and the core, with labels $\boldsymbol{k}_i/A_i$ indicating velocities scaled by the mass numbers $A_1$, $A_2$, and $A_c$. Two Jacobi sets are indicated. In the $(^{15}\mathrm{B}+n)+n$ set (unprimed, Jacobi–Y), $\ell_x$ is the relative momentum of the core–neutron pair $(^{15}\mathrm{B}+n)$ 
with associated energy is $E_{\mathrm{core}\text{-}n}$ and reduced mass $\mu_{2c}$.
$\ell_y$ is the momentum of the second spectator neutron relative to the c.m. of the $^{15}\mathrm{B}+n$ subsystem.  In the $nn$-core set $^{15}\mathrm{B}+(n+n)$ (primed, Jacobi–T), $\ell'_{x}$, $E_{nn}$, and $\mu_{12}$ represent relative momentum, total energy, and reduced mass of the $n+n$ subsystem.  $\ell'_{y}$ is the momentum of the $^{15}\mathrm{B}$ core relative to the $n+n$ c.m. The orbital angular momenta couple to the total as $L=\ell_x\otimes \ell_y=\ell'_x\otimes \ell'_y$. }
\label{fig:diagram}
\end{figure}

In the following, we carry out a detailed study of this exotic situation using advanced theoretical tools based on the Gamow coupled-channel (GCC) framework~\cite{Wang2017,Wang2022,Michel2021}, a three-body formalism that explicitly incorporates continuum effects. In this model, $^{17}\mathrm{B}$ is treated as a composite system of an inert $^{15}\mathrm{B}$ core and two valence neutrons~\cite{Wang2017,Wang2019}. The $^{15}\mathrm{B}$ core’s valence proton is considered a passive spectator, confined to the $0p_{3/2}$ orbital for simplicity. The three-body wave function is formulated in Jacobi coordinates and expanded in a hyperspherical-harmonic basis, which ensures the correct asymptotic behavior of loosely bound systems, see Fig.~\ref{fig:diagram}. For the hyperradial coordinate, we adopt the Berggren ensemble, analogous to that used in the GSM, in order to describe inner structure and decay properties on an equal footing~\cite{Berggren1968,Michel2009,Wang2017}. Finally, antisymmetrization between the deformed core and the valence neutrons is handled via a supersymmetric-transformation technique~\cite{Sparenberg1997}. Further details and parameters of the model used can be found in Appendix~\ref{sec:GCC}.

\subsection{Two-neutron decay width}

\begin{figure}[htb]
\centering
\includegraphics[width=1\columnwidth]{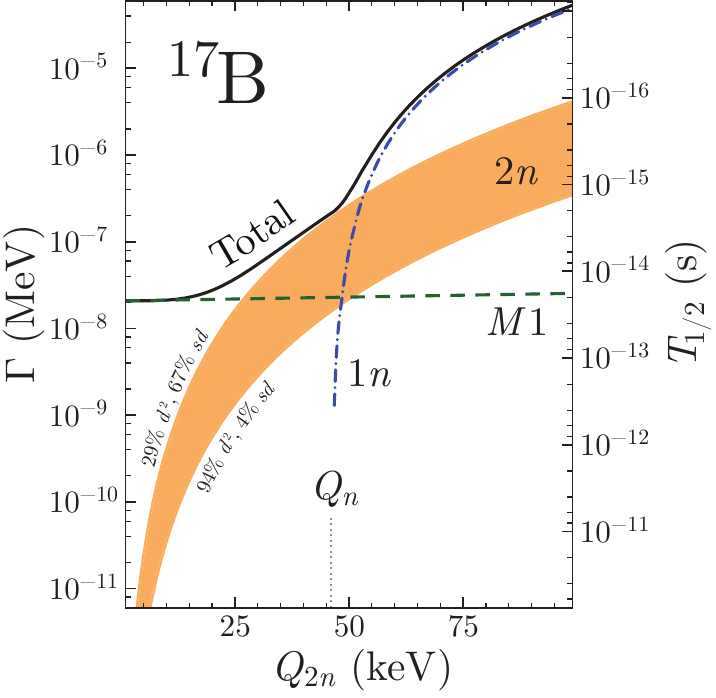}
\caption{Calculated $2n$ (orange band), $1n$ (blue dash-dotted line), $M1$ (green dashed line), and total (black solid line) decay widths (half-lives) from the excited $^{17}$B as a function of $2n$ decay energy $Q_{2n}$. All the particle decays are through the $L = 2$ channel, and spectroscopic factors have been taken into account.}
\label{fig:Decay_width}
\end{figure}

We begin by discussing the competition between $\gamma$ decay and the possible two-neutron decay mechanisms, which depends on the exact position of the $1/2^{-}$ state in $^{17}\mathrm{B}$ relative to the two-neutron threshold; see Fig.~\ref{fig:Decay_width}. 
In this figure, the $M1$ $\gamma$-decay width is shown by the green dashed line and is taken from our structural calculations in Table~\ref{tab:17B_gamma_widths}. 
The orange band represents the two-neutron decay width as a function of the available energy $Q_{2n}$. 
It is shown as a band because we consider several structural limits. 
As discussed in Sec.~\ref{sec:pair}, the two-neutron channel is generally the $S=0$, $L=2$ configuration with two leading components, $d^{2}$ and $sd$. 
The band reflects different assumed weights of these components, as indicated in the figure. 
The blue dot-dash line shows the contribution from direct one-body decay, which becomes possible once the $1/2^{-}$ state lies above the $3^{-}$ level at about 46~keV.
The GCC approach treats one- and two-neutron emission within a unified core plus two-neutron model. 
To incorporate detailed structure, we scale the calculated widths with the appropriate spectroscopic factors discussed in Sec.~\ref{sec:structure}. 
From the figure, it is clear that while the $1/2^{-}$ state remains below the one-neutron threshold (i.e., below the $3^{-}$ resonance in $^{16}\mathrm{B}$), the $M1$ $\gamma$ decay is significant, if not dominant. 
This is enabled by the $L=2$ nature of the two-neutron emission, which is impacted by the reduced phase space at low energy and the centrifugal barrier. In contrast, if the decay proceeded via an $L=0$ channel, the width would increase dramatically. For instance, with $Q_{2n}=10$ keV, our calculated decay width is $7\times10^{-4}$ MeV -- much larger than the predicted $M1$ $\gamma$-decay width. Such a result would contradict the experimental observation of a $\gamma$ ray~\cite{Mozumdar2025Searcha}. Similarly,
once the one-body channel opens, it quickly dominates, even though it proceeds via a $d$ wave. 
The combined width is shown by the solid black line. At higher $Q_{2n}$ the $\gamma$ branch becomes strongly suppressed. Taken together, the results indicate that the $\gamma$ ray is expected to be clearly observable for $Q_{2n}\lesssim 50$~keV and remains consistent with observation for $Q_{2n}$ up to about 100~keV, where the visibility depends on the precise spectroscopic factors, the structure of the state, and experimental conditions.

\begin{figure}[htb]
\centering
\includegraphics[width=1\columnwidth]{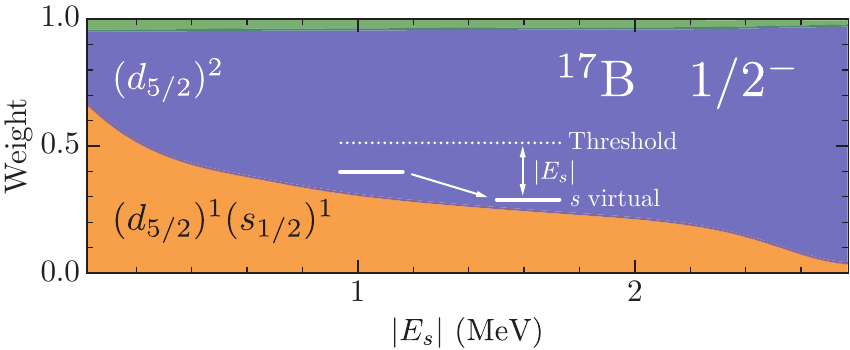}
\caption{Predicted weight of each relative components in the $1/2^-$ excited state of $^{17}$B, plotted as a function of the $s$‑wave virtual-state energy $E_s$ with respect to the threshold. }
\label{fig:Configuration_Evolution}
\end{figure}

As seen from the orange band in Fig.~\ref{fig:Decay_width}, the exact structural configuration of the emitted neutron pair plays a significant role and can easily lead to an order-of-magnitude change of the decay width. 
The SM results quoted above predict mixing of roughly 55\% for $d^{2}$ and 38\% for $sd$. 
However, this configuration, and in particular the structure of the $1/2^{-}$ state, is subject to a significant alignment effect. For the $1/2^{-}$ resonance in $^{17}\mathrm{B}$, which most likely lies between the one- and two-neutron thresholds, our framework describes the intermediate spectrum of $^{16}\mathrm{B}$ in terms of a resonant $d$ orbital together with an antibound (virtual) $s$ state. By fixing the $d_{5/2}$ resonance to the known $^{16}\mathrm{B}$ ground-state energy, which is a $3^{-}$ state, and varying the virtual $s_{1/2}$ energy $E_{s}$ to reflect the influence of an unknown $0^{-}$ or $1^{-}$ state, one finds a dramatic redistribution of configuration weights in the $^{17}\mathrm{B}$ $1/2^{-}$ state; see Fig.~\ref{fig:Configuration_Evolution}. With our baseline parameters, $E_{s}=-0.02$~MeV. Here a small negative value of $E_s$ indicates that the $s$-wave state is virtual (antibound), meaning that it does not form a bound state and lies very close to the threshold.
The $1/2^{-}$ wave function contains 67\% $sd$ and 29\% $d^2$. The decay width in this limit is
$
\Gamma_{2n} = 1.13\times10^{-9}\,\mathrm{MeV}~(E_s = -0.02~\mathrm{MeV}).
$

As the virtual $s_{1/2}$ state is moved further away from the threshold that is, 
as the magnitude $|E_s|$ increases, the $d^2$ component grows to nearly 100\%, 
which changes the two-neutron decay width to
$
\Gamma_{2n} = 5.57\times10^{-11}\,\mathrm{MeV}~(E_s = -2.77~\mathrm{MeV}).
$
Since an additional spectroscopic factor and related changes to the many-body structure must be considered, the true impact may be larger, which helps explain the competition with $\gamma$ emission. 
Furthermore, the significant sensitivity of $\Gamma_{2n}$ to the virtual-state energy and to the components of the wave function would make observation of the two-neutron branch a powerful experimental probe of the internal configuration of the near-threshold $1/2^-$ state.

\subsection{Neutron-neutron angular correlations}
Nucleon–nucleon correlations in the three-body final state provide a sensitive probe of the two-neutron decay channel and quantum configurations involved.
This approach is well established in studies of two-proton emitters~\cite{Pfutzner2023,Zhou2022,Wang2021} and is now being extended to neutron-rich systems. 
The $^{17}\mathrm{B}$ case is especially compelling: the two-neutron decay energy is extremely small (likely similar only to $^{26}\mathrm{O}$~\cite{Kondo2016Nucleus}), and the calculated two-neutron width is comparable to the $M1$ branch, making both modes experimentally accessible. 

\begin{figure}[htb]
\centering
\includegraphics[width=0.85\columnwidth]{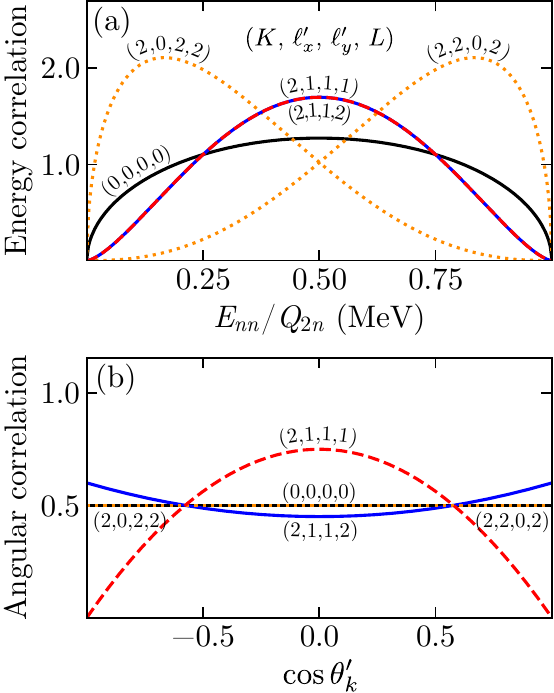}
\caption{(a) Energy and (b) angular correlations for individual configurations in the final state.
Each configuration is labeled by $(K, \ell_x', \ell_y', L)$ in the Jacobi-T coordinate, where $K$ is the hyperspherical quantum number. The definitions of $\ell_x'$, $\ell_y'$, and $L$ are the same as in Fig.~\ref{fig:diagram}.
This plot should be read as a display of wave functions for given Jacobi configurations; the patterns do not depend on whether the $Y$ or $T$ set is used, although the interpretation of the quantum numbers and angles differs between the two.
} 
\label{fig:Correlation_for_individual_configuration}
\end{figure}

In Fig.~\ref{fig:Correlation_for_individual_configuration}, we illustrate how the observed neutron correlations depend on the decay component by showing both energy and angular correlations for individual configurations. 
Each emission configuration is labeled by $(K,\ell'_x,\ell'_y,L)$ in Jacobi-T coordinates, where $K$ is the hyperspherical quantum number (see Fig.~\ref{fig:diagram}). 
At extremely low energies, the structure of the kinetic-energy term in hyperspherical coordinates favors the lowest $K=L$, which is analogous to a centrifugal barrier. 
Thus, $L=0$ two-neutron decays, such as the case of $^{26}\mathrm{O}$, exhibit a simple, symmetric pattern arising from $(K,\ell'_x,\ell'_y,L)=(0,0,0,0)$. 
However, for the present case with $L=2$, several principal configurations contribute. The top panel shows the distribution of the two-neutron energy relative to the total available energy, and the bottom panel shows the distribution of the angle of the neutron pair as defined in Fig.~\ref{fig:diagram}. Although the component-by-component distributions are already distinct, their interference produces additional, significant asymmetries. Therefore the measured $L=2$ angular and energy correlations of $^{17}\mathrm{B}$ can directly constrain the balance of $s_{1/2}$ and $d_{5/2}$ components shaped by near-threshold dynamics and by the influence of resonant and virtual states in $^{16}\mathrm{B}$.

\begin{figure}[htb]
\centering
\includegraphics[width=1\columnwidth]{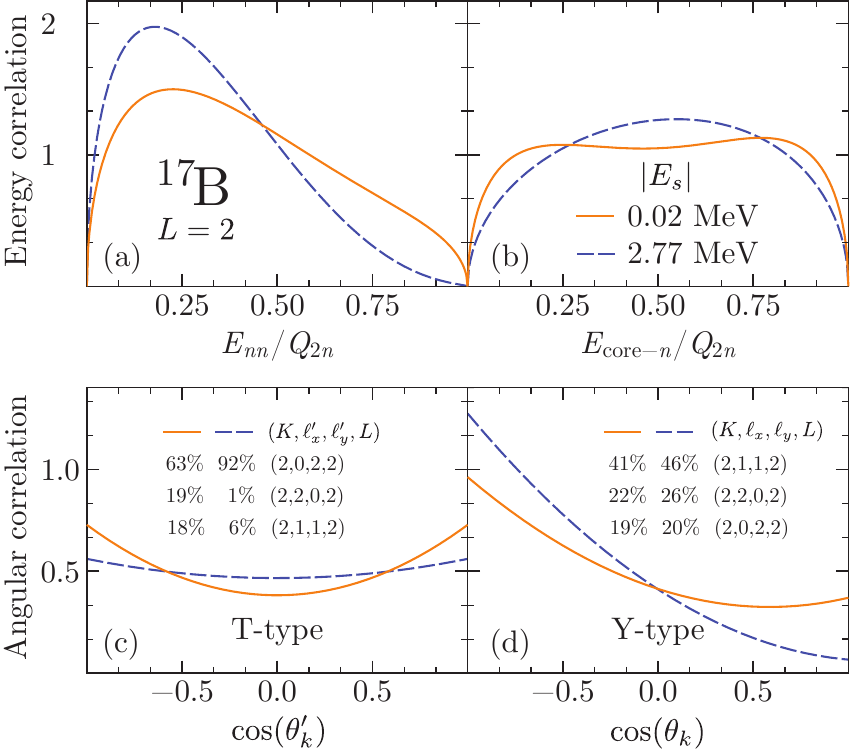}
\caption{(a, b) Predicted energy correlations for the final state of the $L=2$ two‑neutron decay. (c, d) Corresponding angular correlations in the Jacobi-T and -Y coordinates, respectively. The solid curves correspond to an $s$‑wave virtual state positioned extremely close to threshold ($|E_s| = 0.02$ MeV), while dashed curves represent it located farther away from threshold ($|E_s| = 2.77$ MeV). Also listed are the final-state configurations and corresponding weights.}
\label{fig:Configuration_Correlation}
\end{figure}

Our calculation with realistic parameters and full mixing among the components is presented in Fig.~\ref{fig:Configuration_Correlation}. Since the position of the virtual intermediate state ($0^{-}$ and/or $1^{-}$ in $^{16}\mathrm{B}$) is still unknown, we show two limits discussed earlier: a near-threshold virtual state $E_s = -0.02$ MeV (left panels) and an opposite limit 
with $E_s=-2.77$~MeV (right panels).
The results in Fig.~\ref{fig:Configuration_Correlation} show that the observed $nn$ correlation patterns are highly sensitive to the position of the $s$-wave virtual state and, in turn, to the weights of the $sd$ and $d^2$ components. Both cases exhibit asymmetric energy and angular correlations arising from the interference between mixed configurations. 
In particular, when the $1/2^{-}$ resonance decay is dominated by a nearly pure $d^2$ configuration (dashed curves), large opening angles (Fig.~\ref{fig:Configuration_Correlation}d) and high relative kinetic energies (Fig.~\ref{fig:Configuration_Correlation}a) are strongly suppressed. By contrast, an admixture of the spatially symmetric $s_{1/2}$ component produces a more uniform distribution of angles and energy sharing. This contrast makes correlation measurements a promising experimental probe of the internal structure of the near-threshold $1/2^{-}$ resonance. 

The measured nucleon–nucleon correlation reflects the momentum–space configuration of the final state, not the initial intrinsic structure itself. The connection is present but is distorted by the dynamics and final state interactions. 
In Fig.~\ref{fig:Configuration_Correlation}d we show the final-state distribution of angle in the Jacobi-Y coordinate systems. 
For $^{17}\mathrm{B}$, since the recoil kinetic energy of the daughter nucleus is relatively small, at the initial moment of emission the Jacobi-$Y$ configuration $(K,\ell_x,\ell_y,L)$ can be approximately viewed as representing the single-particle configuration, with $\ell_x$ and $\ell_y$ being mean-field–centered angular momenta of the two neutrons. During the decay, however, these configurations evolve. It is found that the initial state with either a nearly pure $d^2$ configuration or a mixed configuration consisting of 67\% $sd$ and 29\% $d^2$ components evolves into a strongly modified final state. In both cases, the $d^2$ component nearly vanishes in the final state because the high centrifugal barrier associated with the $d$ wave suppresses tunneling of Jacobi configurations with $\ell_x=\ell_y=2$ at extremely low energies. 
Meanwhile, more than 40\% of the final configuration is represented by the $p^2$ component with $\ell_x=\ell_y=1$, which emerges dynamically in the tunneling process due to the overall attractive nucleon–nucleon interaction that favors di-neutron formation so that, in the emission process, both neutrons are subject to comparable centrifugal barriers.

\begin{figure}[htb]
\centering
\includegraphics[width=1\columnwidth]{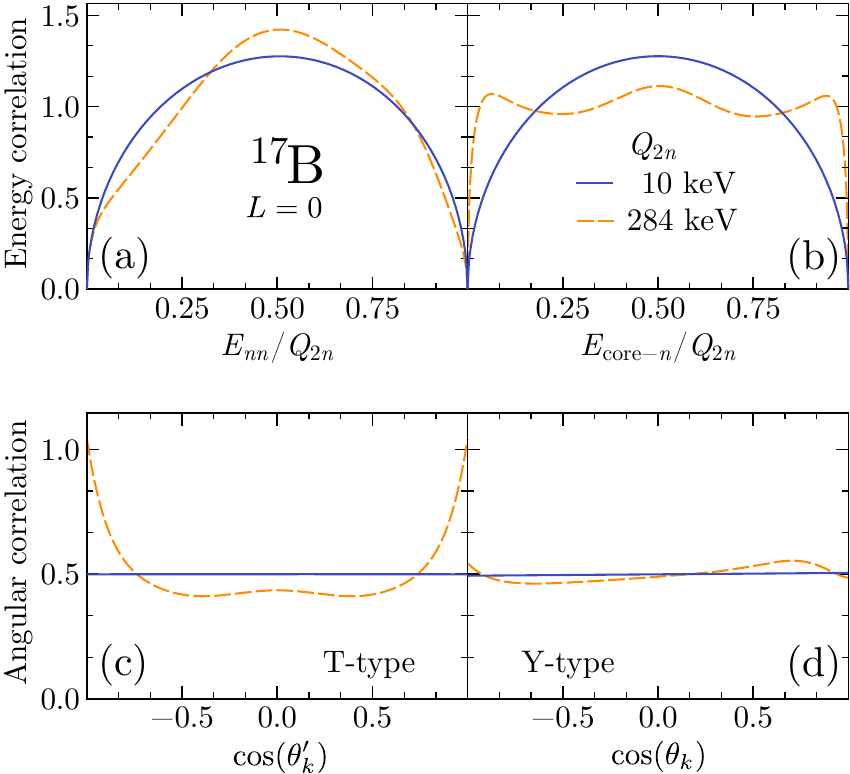}
\caption{Predicted energy (a,b) and angular (c,d) correlations of $L = 0$ two-neutron ($2n$) decay from the excited state of $^{17}$B as different decay energies $Q_{2n}$. The solid and dashed lines represent the results for $Q_{2n} = 10$ and 284 keV, respectively. }
\label{fig:Correlation_17B_threshold}
\end{figure}

For completeness, it is useful to address the $L=0$ two-neutron emission and the corresponding angular correlations. This is practically relevant because a higher-lying $3/2^{-}$ state in $^{17}\mathrm{B}$, which is fully unbound, is predicted by the SM, and neutrons from that energy region have been targeted experimentally~\cite{Mozumdar2025Searcha}. This SM–predicted second $3/2^{-}$ state (a broad resonance) in $^{17}\mathrm{B}$ is expected above the one-neutron threshold, and its decay should be dominated by $L=0$ emission to the $3/2^{-}$ state in $^{15}\mathrm{B}$.
This leads to markedly different three-body correlations (see Fig.~\ref{fig:Correlation_17B_threshold}). When the $3/2^{-}$ state lies very close to the two-neutron threshold (analogous to $^{26}\mathrm{O}$), only $s$-wave decay is present; other partial waves are strongly suppressed by the centrifugal barrier and reduced phase space, and both energy and angular correlations are essentially featureless and nearly uniform. As the decay energy rises toward $\sim 300$~keV, structures begin to appear. In particular, the $E_{\mathrm{core}\text{-}n}$ spectrum develops distinct peaks, indicating the onset of competition between sequential one-neutron emission and direct two-neutron decay. Since the $L=0$ decay is dominated by the $s$ wave, the neutron–neutron correlation in the final state remains largely symmetric.

\section{Conclusion}
In this work, we study a unique excited state in $^{17}\mathrm{B}$. 
A $\gamma$ ray of 1640~keV has been reported in its decay to the ground state of $^{17}\mathrm{B}$, while the best estimates place the two-neutron threshold at 1380(210) keV, \cite{and2012Ame2012} and the one-neutron threshold only a few tens of keV higher, with several resonances observed in $^{16}\mathrm{B}$. 
Our analysis of reaction mechanisms, $\gamma$ decay, and SM systematics all point to an unambiguous $1/2^{-}$ assignment for this state. 
This represents a novel situation in which $\gamma$ decay competes with neutron emission. We identify the key ingredients that make such competition possible: $L=2$ emission of the neutron pair and a complex structural interplay between $s$- and $d$-wave single-particle components in boron isotopes that is strongly influenced by weak binding and continuum coupling.
We carried out a detailed study using both a traditional SM and the GSM, focusing on the structure of the emitted pair, which is predominantly a spin-singlet with $L=2$ and two leading configurations, $s_{1/2}d_{5/2}$ and $(d_{5/2})^2$, in a mean-field basis. 
This configuration mixing is driven not only by the distinctive structure of boron isotopes discussed in the literature, but is also highly sensitive to the position of $0^{-}$ or $1^{-}$ resonances in $^{16}\mathrm{B}$ that couple the one-body $s$-wave neutron continuum to the decaying $1/2^{-}$ state in $^{17}\mathrm{B}$.

We complemented these structural studies with an advanced treatment of the two-neutron decay using the GCC framework. 
This provides a quantitative assessment of the decay channels and widths and confirms potential competition between two-neutron emission and $\gamma$ decay over a substantial range of near-threshold energies. 
The resulting width is highly sensitive to configuration mixing and to the position of the $s$-wave–coupled resonant structures in $^{16}\mathrm{B}$.
The observation and analysis of neutron–neutron correlations from this decay are of paramount interest for elucidating the internal structure, neutron configurations, and the interference between configurations and between sequential and direct decay paths. Using the same framework, we explored the expected correlation patterns as functions of the two-neutron separation energy, configuration mixing, and decay mechanism. 
We addressed the $1/2^{-}$ state of primary interest, which involves $L=2$ decay, and contrasted it with the $L=0$ limit expected for decay from a higher-lying $3/2^{-}$ state.

This work is timely: the case has become experimentally accessible only recently, and it sits at the interface of nuclear structure, reaction dynamics, and open-quantum-system physics. The $^{17}\mathrm{B}$ resonance provides a realistic and, to our knowledge, first-of-its-kind opportunity to apply advanced nuclear theory at the edge of stability, where continuum coupling, near-threshold dynamics, and configuration mixing all play decisive roles.

\begin{acknowledgments}
The authors gratefully acknowledge the support of GANIL through its Visiting Scholar Program, which made this work possible. We also thank Olivier Sorlin and Nicolas Michel for stimulating discussions.
This material is based upon work supported by the U.S. Department of Energy, Office of Science, Office of Nuclear Physics under Award No. DE-SC0009883; the National Key Research and Development Program of China (MOST 2023YFA1606404 and MOST 2022YFA1602303); the National Natural Science Foundation of China under Contract No.\,12347106, No.\,12147101, No.12447122, and the China Postdoctoral Science Foundation under Contract No.\,2024M760489. Part of this work was performed using computing resources of CRIANN (Normandy, France). 
\end{acknowledgments}

\noindent\makebox[\linewidth]{\rule{0.3\linewidth}{0.6pt}}


\appendix
\section{Details of GSM approach\label{sec:GSM}}
In this work, the GSM with MBPT are used~\cite{Hu2020,Xu2023}.
The chiral interaction EM1.8/2.0~\cite{Hebeler2011} is employed, which can globally reproduce nuclear binding energies~\cite{Stroberg2021,Miyagi2022}. 
To reproduce the separation energies of $^{16}\mathrm{B}$ and $^{17}\mathrm{B}$, which are important for the structure and decay properties of near-threshold states, we slightly adjust the low-energy constant of the tensor force, $C_{{}^3S_1-{}^3D_1}$, from $-0.147167$ to $-0.125092$ (in units of $10^{4}\,\mathrm{GeV}^{-2}$). 

The nucleus $^{16}\mathrm{O}$ serves as the reference state for Gamow Hartree–Fock  calculations~\cite{Hagen2006,Zhang2023} and as the GSM core. 
The $0p_{3/2}$ and $0p_{1/2}$ orbitals are chosen as the proton hole space, while the $sd$ orbitals are chosen as the neutron particle space. The Gamow Hartree-Fock calculations yield resonant $\nu 1s_{1/2}$ and $\nu 0d_{3/2}$, which are treated in the complex-$k$ basis to include continuum effects. The rest orbits are treated in discrete, real-energy Hartree–Fock basis. The complex-momentum contours for $\nu s_{1/2}$ and $\nu d_{3/2}$ are $k=0 \rightarrow 0.55-0.10i \rightarrow 1.10 \rightarrow 4~\text{fm}^{-1}$ and $k=0 \rightarrow 0.44-0.20i \rightarrow 0.88 \rightarrow 4~\text{fm}^{-1}$, respectively. Each partial wave contour is discretized with 35 scattering states.
Prior to renormalization, we use a harmonic–oscillator basis to evaluate the interaction matrix elements, with $\hbar\omega=16$~MeV, 13 major shells ($e=2n+\ell\le e_{\max}=12$), and $e_{3\max}=e_1+e_2+e_3\le 12$ for the three–nucleon force. 
The valence-space effective Hamiltonian is renormalized from the full-space Hamiltonian via MBPT, following the procedure in~\cite{Xu2023}.
In the GSM calculation, we restrict at most two valence particles to occupy scattering states, which is sufficient for convergence~\cite{Hu2020}.

\section{Details of GCC approach\label{sec:GCC}}
Within the GCC approach in this work we employ the Minnesota potential with its original parametrization \cite{Thompson1977} to describe the interaction between the valence nucleons. For the core–valence channel, a Woods–Saxon central potential augmented by a spin–orbit term is used \cite{Wang2017}. The spin–orbit strength is set to $V_{\rm s.o.}=30\,$MeV, the diffuseness to $a=0.7\,$fm, and the central and spin–orbit radii to $R_0=3.07\,$fm and $R_{\rm s.o.}=3.07\,$fm respectively. Because the low-lying states of $^{17}$B lie near threshold and are highly sensitive to continuum coupling, we fine-tune the central depth $V_0$ to reproduce the one-neutron separation energy $Q_n=46\,$keV of $^{16}$B. Furthermore, we introduce an effective three-body force, represented by a hyperradial Woods–Saxon form without spin–orbit contribution. Its diffuseness and radius are fixed at $a^{3N}=0.7\,$fm and $R_0^{3N}=5\,$fm, respectively, while the depth $V_0^{3b}$ is adjusted to reproduce the empirical two-neutron separation energy $Q_{2n}$ of the system.

Calculations are performed within a model space defined by $\max(\ell_x,\ell_y)\leq11$ and a maximal hyperspherical quantum number $K_{\max}=20$ \cite{Wang2017}. To study resonances, the Berggren basis is utilized for $K_{\max} \leq 4$ channels, complemented by a harmonic oscillator basis with $b = 1.75$ fm and $N_{\max} = 40$ for higher-angular-momentum channels. The complex-momentum contour for the Berggren ensemble spans $k:0\to0.3-0.1i\to0.5\to1.2\to6\,$fm$^{-1}$, discretized into 80 scattering states per segment to ensure convergence of continuum contributions.
\bibliography{bib17B,17B}

\end{document}